\documentclass[aps,pra,reprint,superscriptaddress,showpacs]{revtex4-1}
 \input{header}

\begin{document}


\title{Evaporative cooling of cold atoms at surfaces}


\author{J.M\"arkle}
\affiliation{CQ Center for Collective Quantum Phenomena and their Applications in LISA+, Physikalisches Institut der Universität Tübingen, Auf der Morgenstelle 14, D-72076 Tübingen, Germany}
\author{A.J.Allen}
\affiliation{Joint Quantum Centre (JQC) Durham-Newcastle, School of Mathematics and Statistics, Newcastle University,
Newcastle upon Tyne, NE1 7RU, England, UK}
\author{P.Federsel}
\author{B.Jetter}
\affiliation{CQ Center for Collective Quantum Phenomena and their Applications in LISA+, Physikalisches Institut der Universität Tübingen, Auf der Morgenstelle 14, D-72076 Tübingen, Germany}
\author{A.G\"unther}
\author{J.Fort\'agh}
\affiliation{CQ Center for Collective Quantum Phenomena and their Applications in LISA+, Physikalisches Institut der Universität Tübingen, Auf der Morgenstelle 14, D-72076 Tübingen, Germany}
\author{N.P.Proukakis}
\affiliation{Joint Quantum Centre (JQC) Durham-Newcastle, School of Mathematics and Statistics, Newcastle University,
Newcastle upon Tyne, NE1 7RU, England, UK}
\author{T.E.Judd}
\affiliation{CQ Center for Collective Quantum Phenomena and their Applications in LISA+, Physikalisches Institut der Universität Tübingen, Auf der Morgenstelle 14, D-72076 Tübingen, Germany}

\date{\today}

\begin{abstract}
We theoretically investigate the evaporative cooling of cold rubidium atoms that are brought close to a solid surface. The dynamics of the atom cloud are described by coupling a dissipative Gross-Pitaevskii equation for the condensate with a quantum Boltzmann description of the thermal cloud (the Zaremba-Nikuni-Griffin method). We have also performed experiments to allow for a detailed comparison with this model and find that it can capture the key physics of this system provided the full collisional dynamics of the thermal cloud are included. In addition, we suggest how to optimize surface cooling to obtain the purest and largest condensates.
\end{abstract}



\maketitle

\section{INTRODUCTION}
Since the advent of microchip traps for cold atoms \cite{PhysRevLett.81.5310,Denschlag1999,hinds1999magnetic,Folman2000,Ott2001,Hansel2001,PhysRevA.66.041604,PhysRevA.71.063619,PhysRevA.66.051401,Hinds_BEC_Chip,Fortagh2007,whitlock2008}, interest in developing quantum hybrid systems, which exploit the long coherence times of Bose-Einstein condensates with the flexibility of modern micro- and nanoelectronics, continues to grow. There is potential to use such systems as quantum memory devices \cite{PhysRevA.69.062320,Treutlein2007,Treutlein2010}, precision measurement devices \cite{Wang2006,gierling2011cold,schneeweiss2012dispersion,1367-2630-15-7-073009} and even rewritable electronic systems \cite{Judd2010}. More recently, there have been proposals to use cold atoms to cool nanoscaled solid objects \cite{Chang2010,weiss2013}; ion cooling using neutral atoms has already been demonstrated  \cite{Zipkes2010, Schmid2010}.

As a result of these experimental advances, a need has grown to develop theoretical tools that can describe hybrid devices at finite temperatures. A range of methods have been previously developed for describing finite temperature cold gases in isolation \cite{Proukakis2008,proukakis2013quantum,ZNGbook,blakie_bradley_08,Griffin1996,svistunov,Stoof1997,Gardiner1998,proukakis_burnett_98,stoof_99,Zarmeba1999,Walser1999,davis_qk2000,Bijlsma2000,stoof_bijlsma_01,Davis2001,Gardiner2002,Blakie2005,Cockburn2009,gasenzer_09,Zhu2013}, but none have been used in the context of hybrid devices.

A challenging test of a finite-temperature method is the problem of evaporative cooling when atoms are brought close to a solid surface. This is, in fact, a rather common experiment in the atom chip community, where surface losses are frequently used to calibrate the position of the surface. Such experiments lead to non-trivial loss curves \cite{hindslosses,Kasch2010} and are known to be an efficient route to Bose-Einstein condensation \cite{Harber2003}. 
Previous work on free-space evaporative cooling (see, e.g. Ref.~\cite{Ketterle1996} for an early review) has typically been based on the classical ergodic Boltzmann equation \cite{Luiten1996,Wu1996,holland_evapcooling}, extended to include rate equations for the losses \cite{yamashita_evapcooling}, 
or on phase-space methods \cite{drummond_evapcooling}. 
Condensate growth has also been studied by sudden truncation of the thermal distribution in the ergodic approximation \cite{davis_qk2000,Bijlsma2000,davis_gardiner_02,hugbart_retter_07}, or through a dynamical quench \cite{stoof_bijlsma_01,Duine2001,proukakis_schmiedmayer_06,weiler_neely_08,damski_zurek_10,garrett_2011}.
However, the surface cooling problem, 
which requires a detailed description of the atomic collisional processes for both the condensate and the thermal cloud
has not yet been theoretically studied, either qualitatively or quantitatively. In addition, further experimental work is required to provide benchmarks for such theories. 
%

The aim of this paper is to show that the key physics of surface evaporative cooling may be captured using the Zaremba-Nikuni-Griffin (ZNG) model \cite{Zarmeba1999,ZNGbook} in its full dynamical non-equilibrium implementation, extending beyond ergodicity \cite{Jackson2002}. The ZNG kinetic model accounts for full collisional redistribution between the condensate and the thermal cloud, taking Bose enhancement and BEC growth into account. It has been previously applied successfully to a diverse range of problems \cite{Jackson2001,1367-2630-5-1-388,jackson_zaremba_02c,jackson_proukakis_07,Jackson2009,Allen2013}.
To demonstrate the applicability of this model to the problem of surface evaporative cooling, we also present previously unreported experimental results based on $^{87}\text{Rb}$ and a silicon surface,
revealing consistency between theory and experiment. In spite of neglecting fluctuations around the phase transition, this method appears to be able to describe condensate growth and the non-trivial atom loss curves observed in experiments. At the end of this study, we show how to optimize the surface cooling of a cold cloud to obtain the purest or largest condensates.

%

\section{METHODS}
We begin by briefly reviewing the ZNG formalism \cite{Zarmeba1999,ZNGbook} for describing a cold cloud of $N$ atoms at finite temperature. In this model, thermal excitations are treated semi-classically within the Hartree-Fock and Popov-approximations \cite{Griffin1996}. This leads to a generalised Gross-Pitaevskii equation (GPE) for atoms of mass $m$, describing the time evolution of the condensate wavefunction $\Psi(\textbf{r},t)$ in an external potential $V(\textbf{r})$
\begin{align}
	i\hbar &\frac{\partial\Psi}{\partial t}=\left(-\frac{\hbar^2\nabla^2}{2m} + V + gn_c +2g\tilde{n}-iR\right)\Psi, \label{GPE}
	\intertext{which is coupled to a quantum Boltzmann equation for the thermal atoms}
	&\frac{\partial f}{\partial t}+\frac{\textbf{p}}{m}\cdot\nabla f-\nabla U\cdot\nabla_{\textbf{p}} f = C_{12}[f,\Psi]+C_{22}[f]. \label{QB}
\end{align}
Here, $f(\textbf{r},\textbf{p},t)$ is the phase space density of the thermal cloud, $n_c(\textbf{r},t)=|\Psi(\textbf{r},t)|^2$ is the condensate spatial density, $\tilde{n}(\textbf{r},t)=\int (d\textbf{p}/h^3) f(\textbf{p},\textbf{r},t)$ is the thermal cloud spatial density, $iR$ is a source term that leads to loss or gain of condensate atoms, $\textbf{p}$ is the atomic momentum vector, $U(\textbf{r},t)= V(\textbf{r}) + 2g(n_c(\textbf{r},t) + \tilde{n}(\textbf{r},t))$ is the effective potential experienced by the thermal atoms, combining the external and interaction potentials, and $C_{12}$ and $C_{22}$ are the collision integrals. The strength of the atomic interactions in the condensate is given by the coupling constant $g=4\pi\hbar^2a/m$, where $a$ is the $s$-wave scattering length ($\approx5.4\:$nm for $^{87}$Rb). The other symbols have their usual meaning with $\nabla$, $\nabla_{\textbf{p}}$ representing the three-dimensional derivatives with respect to space and   momentum. 

The condensate density is normalized to the current number of condensate atoms, $N_c$, and the thermal cloud density $\tilde{n}(\textbf{r},t)$ is obtained by integrating $f(\textbf{r},\textbf{p},t)$ over momentum space. 
The two densities appear not only in Eq.\ (\ref{GPE}) for the condensate, but also in the expression for the effective potential $U(\textbf{r},t)$. This leads to a mean-field coupling between the condensate and the thermal cloud. In addition to this mean-field coupling, atoms in the thermal cloud can scatter from one another ($C_{22}$~collisions), and atoms can scatter into or out of the condensate ($C_{12}$~collisions). These collisions are calculated via the collision integrals
\begin{align}
C_{22}[f]=&\frac{2g^2}{(2\pi)^5 h^7}\int d \textbf{p}_2 d\textbf{p}_3 d\textbf{p}_4\delta(\textbf{p}+\textbf{p}_2-\textbf{p}_3-\textbf{p}_4) \notag 
\\ &\times\delta(e + e_2 -e_3 -e_4)	\notag 
\\ &\times [(1+f)(1+f_2)f_3 f_4-f f_2 (1+f_3)(1+f_4)] \label{Eq:C22}
\\C_{12}[f,\Psi]=&\frac{2g^2n_c}{(2\pi)^2h^4}\int d\textbf{p}_2 d\textbf{p}_3 d\textbf{p}_4 \delta(m \textbf{v}_c + \textbf{p}_2 - \textbf{p}_3 -\textbf{p}_4) \notag
\\ &\times\delta(e_c + e_2 - e_3 - e_4) \notag
\\ &\times[\delta(\textbf{p}-\textbf{p}_2) -\delta(\textbf{p}-\textbf{p}_3) -\delta(\textbf{p}-\textbf{p}_4)] \notag
\\ &\times[(1+f_2)f_3f_4 - f_2(1+f_3)(1+f_4)] \label{Eq:C12}
\end{align}
where $f_i\equiv f(\textbf{p}_i,\textbf{r},t)$ and $e_i=\textbf{p}_i^2/2m +U(\textbf{r},t)$. They consider all two-body scattering events; $\delta$-functions ensure momentum and energy conservation.
To accurately describe the energies involved in a $C_{12}$ collision, it is necessary to include the local condensate energy $e_c=m\textbf{v}_c(\textbf{r})^2/2+\mu_c(\textbf{r})$, where $\mu_c(\textbf{r})$ is the chemical potential and $\textbf{v}_c(\textbf{r})$ is the local condensate velocity \cite{pethick2002}. If an atom leaves the condensate or goes into the condensate because of a scattering event, the normalization of the GPE needs to be changed accordingly using the non-hermitian source term $-iR(\textbf{r},t)$ in the GPE which is defined with the $C_{12}$ collision integral as:
\begin{align}
R(\textbf{r},t)=\frac{\hbar}{2n_c}\int\frac{d\textbf{p}}{(2\pi\hbar)^3}\,C_{12}[f(\textbf{r},\textbf{p},t),\Psi(\textbf{r},t)].
\end{align} 
We now discuss the specifics of applying the ZNG theory to the problem of surface evaporative cooling.

\begin{figure}
\begin{center}
\includegraphics{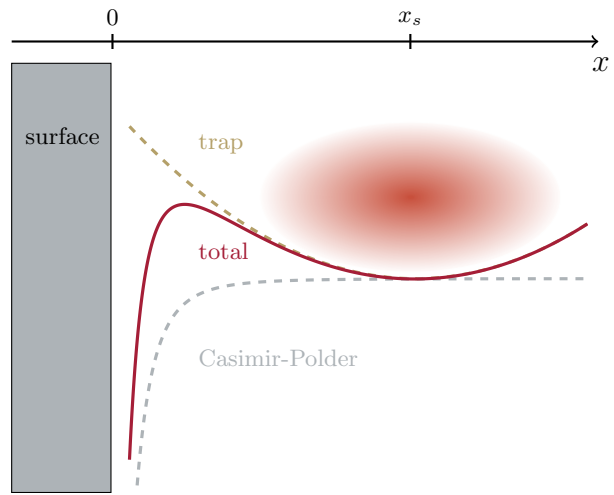}
\caption{(Color online) Schematic diagram (not to scale) of the system showing the total potential $V(x,y=0,z=0)$ (red solid curve) for atoms in a harmonic trap in the vicinity of a surface. For short distances the Casimir-Polder potential (gray dashed curve) dominates. Far away from the surface the atoms only see the trapping potential (gold dashed curve). In between, the Casimir-Polder potential leads to an opening of the trap. Solid rectangle indicates the surface and the colored oval indicates the atom cloud. Black arrow shows the $x$-axis with $x_s$ the distance between the trap center and the surface.}
\label{fig:pot}
\end{center}
\end{figure}
\section{IMPLEMENTATION}

In our system the external potential is given by a combination of
the trapping potential and the potential due to the surface, and takes the form
\begin{align}
\label{eq:potential}
V(\textbf{r})=\frac{1}{2}m\omega_x^2 (x-x_s)^2 +\frac{1}{2}m\omega_y^2 y^2 +\frac{1}{2}m\omega_z^2 z^2 + V_{CP}(x).
\end{align}
The first three terms represent a harmonic trapping potential, centered at $x=x_s$, where $x_s$ is the distance of the trap center [Fig.\ \ref{fig:pot}] from the surface, defined as the $x=0$ plane.
Trap frequencies are $\omega_{i=x,y,z}$ in the $x$, $y$ and $z$ directions respectively. The Casimir-Polder potential $V_{CP}(x)$ describes the interaction between an atom and the surface, approximated using a single-correction function \cite{Shimizu2001}
\begin{align}
V_{CP}(x)=-\frac{C_4}{x^3\cdot(x+\frac{3\lambda}{2\pi ^2})}. \label{EQ:SCF}
\end{align}
Here, $\lambda=780\:$nm is the effective atomic transition wavelength for $^{87}\text{Rb}$ and $C_4$ is a material constant of the form
\begin{align}
	C_4=\frac{3\hbar c \alpha}{32\pi^2\epsilon_0}\left(\frac{\epsilon -1}{\epsilon +1}\right)\Phi(\epsilon),
\end{align}
with $\epsilon$ the relative permittivity of the surface, $\alpha$ the static polarizability of the atom, and $\Phi(\epsilon)$ a dimensionless constant for the surface \cite{Zong-Chao1997}. The other symbols have their usual meaning. For a silicon surface and $^{87}\text{Rb}$, $C_4=1.22e-55\:$J$\text{m}^4$ \cite{Lin2004}. 

The potentials are sketched in Fig.\ \ref{fig:pot}. Far away from the surface the atoms only see the trapping potential whereas close to the surface the attractive Casimir-Polder potential dominates. In the intermediate regime the Casimir-Polder potential leads to an opening of the trap [solid red curve] and atoms are lost to the surface.

The generalized Gross-Pitaevskii equation (Eq.\ (\ref{GPE})) can be solved using a split-step method \cite{Kuhn2011}, as outlined in Ref.\ \cite{ZNGbook}. This can be parallelised using the FFTW package \cite{FFTW05} combined with the message-passing-interface (MPI) \cite{gabriel04:_open_mpi}.

In the full ZNG numerical implementation \cite{Jackson2002} the quantum Boltzmann equation (Eq.\ (\ref{QB})) is iterated in time by a direct-simulation-Monte-Carlo (DSMC) \cite{Gallis20094532} approach, in which a swarm of test particles models the distribution function $f(\textbf{r},\textbf{p},t)$. The collision integrals in Eq.\ (\ref{Eq:C22}) and Eq.\ (\ref{Eq:C12}) are then replaced by collision probabilities for each test particle. The test particles are binned into collision cells to determine possible collision partners. Because the density of the atom cloud can vary considerably, we use an adaptive cartesian grid in real space as outlined in \cite{Wade2011}, while keeping a global time step.

Our initial state is a thermal cloud in equilibrium with a temperature $T$. This state is calculated using self-consistent Hartree-Fock as outlined in \cite{Williams2001}. 
In addition to the thermal cloud, the initial state requires a small condensate ``seed'' to allow for $C_{12}$ collisions, and hence condensate growth; the number of atoms in the seed is obtained using the Bose-Einstein distribution, assuming $\mu_c=0$ \cite{Bijlsma2000}.

Interactions between the surface and the atoms are modeled by calculating the single-correction function (Eq. (\ref{EQ:SCF})) for the generalized Gross-Pitaevskii equation (Eq. (\ref{GPE})) and combining it with a linear imaginary potential to remove condensate atoms, effective from the position where the trap opens \cite{Judd2008}. In addition, we annihilate test particles that are beyond this opening point, resulting in an atom loss for the thermal cloud.
These two processes lead to a reduction in the total atom number in the system.
\section{RESULTS}
Having set up our computational model, we now employ it to study surface evaporative cooling. We show the results of simulations for two different geometries. In Subsection A, we directly compare theory with experiment to examine the extent to which the model captures the important physical processes. We then go on to consider a simpler model system in Subsection B with a view to optimizing parameters to create the purest or largest condensates.

The experiments were performed using the apparatus described in \cite{gierling2011cold}. Clouds of $^{87}$Rb atoms were loaded into an atom chip trap with frequencies $\omega_x = 2\pi \times 16\:\text{rads}\:\text{s}^{-1}$ in the axial direction and $\omega_y = \omega_z = 2\pi \times 85\,\text{rads}\,\text{s}^{-1}$ in the radial direction. The cloud was initially prepared with the trap center at a distance $x_s\approx 135\:\mu$m from a silicon surface, defined as the $x=0$ plane. At this point there was negligible overlap between the cloud and the surface. The cloud was then transported along the $x$-axis at a variable speed to a variable distance, $x_s$, from the surface and held for a variable hold time. In order to measure the remaining atom number, $N$, the cloud was swiftly brought back to its initial position, after which we performed time-of-flight measurements and CCD imaging.

\subsection{Loss curves}
\subsubsection{Time series}
We begin by initially considering atom loss curves as a function of time when the cloud is brought into overlap with the surface. In the experiments the cloud was transported  to the surface in $1\:$s and held stationary at a final hold point for up to $2.5\:$s. Three hold points were considered: $x_s \approx 14\:\mu$m, $29\:\mu$m and $72\:\mu$m. These were estimated from the point where the trap completely opened and all the atoms were lost to the surface. Reference measurements revealed that temperature-related drifts could shift the position of the surface by up to $10\:\mu$m, hence the given values for $x_s$ are approximate; this is the dominant source of error. The initial cloud temperatures were $130\:$nK for $x_s\approx 14\:\mu$m and $x_s\approx29\:\mu$m, and $140\:$nK for $x_s\approx72\:\mu$m. These temperatures are slightly above the critical temperature for condensation, $T_c$, for an ideal gas \footnote{The critical temperature may be estimated with $T_c=0.94\hbar(\omega_x\omega_y\omega_z)^{1/3}N^{1/3}$ \cite{pethick2002}}.

\begin{figure}
\includegraphics{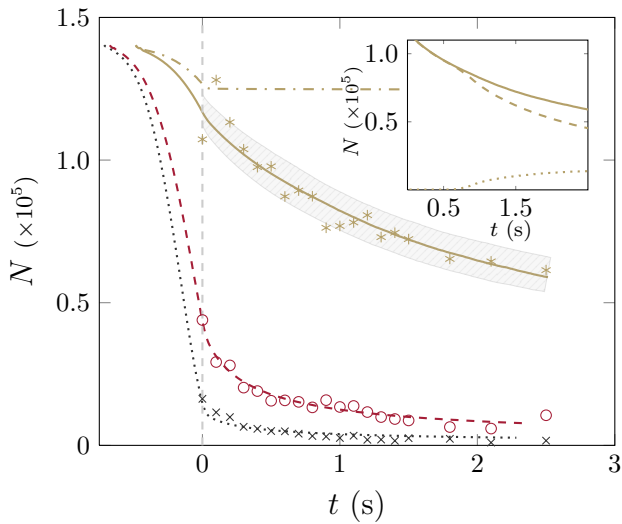}
\caption{(Color online) Total atom number, $N$, against time $t$ for three different trap-surface separations: $x_s=$ $68.0\:\mu$m (gold solid curve), $30.0\:\mu$m (red dashed curve) and $15.0\:\mu$m (black dotted curve). Points correspond to experimental data and the curves correspond to simulations. The dot-dash gold curve shows a simulation for $x_s=68.0\:\mu$m without collisions i.e.\ $C_{12} = C_{22} = 0$. The gray vertical dashed line marks the point when the atom cloud reaches its final hold position at $t=0$. Gray hashed area shows the shift of the curve when the surface position is varied by $\pm 2.5\:\mu$m.
The inset shows a breakdown of the cloud atom numbers against time for $x_s=68.0\:\mu$m from the point when the cloud reaches its holding position. Solid curve shows the total atom number, dashed curve corresponds to thermal atoms, and the dotted curve to the condensate atom number.}
\label{fig:lossRates}
\end{figure}

We performed the simulations using these experimental parameters \footnote{We used a slightly different start position in the simulation of $x$=100\:$\mu$m as a numerical convenience. There was still very little atom-surface overlap at this point so this modification does not affect the results beyond shifting the absolute times.}. We plot the theoretical and experimental atom numbers against time in Fig.\ \ref{fig:lossRates} (a ``time-series''). We consider the time $t=0$ to be the point when the cloud reaches its final hold position indicated by the gray vertical dashed line. Since the absolute surface position may vary due to drifts, we performed a range of simulations with varying $x_s$ to obtain the best fit. In this sense the simulations served as a calibration tool: for the $x_s\approx 14\:\mu$m, $x_s\approx 29\:\mu$m and $x_s\approx 72\:\mu$m curves, the best fits were obtained with a simulated cloud-surface separation of $15.0\:\mu$m, $30.0\:\mu$m, and $68.0\:\mu$m respectively, well within the experimental uncertainties.
Figure \ref{fig:lossRates} shows the evolution of the total number of atoms, $N$, remaining in the cloud during the course of the simulation with curves corresponding to numerical results, and points to experimental data.

The gold solid curve and gold star points are for the $x_s=68.0\:\mu$m hold point, the red dashed curve and red open circles are for the $x_s=30.0\:\mu$m hold point, and the black dotted curve and black crosses are for the $x_s=15.0\:\mu$m hold point. To give an idea of how the surface position affects the remaining atom number we vary the surface position by $\pm 2.5\:\mu$m for the $68.0\:\mu$m curve, shown as the gray hashed area in Fig.\ \ref{fig:lossRates}. 

For all values of $x_s$ we observe a non-trivial loss curve; the loss rates increase to a maximum as the cloud is brought to the surface. Once the cloud reaches its final position, the losses swiftly reduce. The transfer between these regimes is especially pronounced for the red dashed and black dotted curves, where most of the atoms are lost, and a sharp ``elbow'' is observed.
\begin{figure*}[t]
\includegraphics{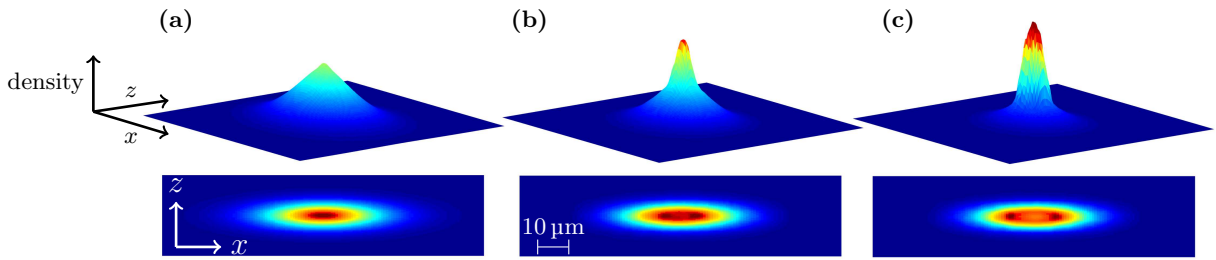}
\caption{\label{fig:densities}(Color online) Density profiles of a cooling ZNG gas at (a) the beginning of the simulation before transport, (b) time $t=0.6\:$s hold time and (c) $t=2.25\:$s hold time.  The upper panels show the full cloud density, integrated along the $y$-direction and the lower panels show cross-sections of the thermal cloud density through the $y=0$ plane. Arrows indicate axes and bars indicate scale. In the upper panels, the $z$-direction has been stretched by a factor of 4 to improve clarity.}
\label{fig:densProf}
\end{figure*}
The initial fast losses occur because atoms are forced over the trap edge during transport. On reaching the final position, the atoms with sufficient kinetic energy in the $x$-direction [see Fig.\ \ref{fig:pot}] are lost within one trap period. After this, a slow loss of atoms still continues because the gas re-thermalizes due to the collisions. The importance of describing the re-thermalization correctly is shown by the gold dot-dash curve in Fig.\ \ref{fig:lossRates}, which repeats the simulation for $x_s=68.0\:\mu$m without any collisions i.e. setting $C_{12}=C_{22}=0$ in the solution of Eq.\ (\ref{QB}) so $iR = 0$ in Eq. (\ref{GPE}). We see that once all the losses due to transport have occurred, no further losses take place and the deviation of this curve from the collisional simulation and experimental data points is stark. The overlap between theory and experiment suggests that losses due to three-body recombination of atoms should be small. Calculations using Refs. \cite{Burt1997} and \cite{Esry1999} and the initial thermal density return three-body loss rates of no more than 500 atoms per second, confirming that this is the case.

Surface evaporative cooling has already been experimentally demonstrated \cite{Harber2003} as an effective route to Bose-Einstein condensation,
with condensate formation clearly observed in those experiments.
Our theoretical scheme also predicts the gradual formation of a condensate; this is shown in the inset of Fig.\ \ref{fig:lossRates} for the $x_s=68.0$\:$\mu$m simulation, depicting the characteristic growth curve \cite{miesner_stamper-kurn_98,davis_qk2000,Bijlsma2000,kohl_davis_02,proukakis_schmiedmayer_06,hugbart_retter_07,garrett_2011}.
The surface removes the hottest atoms from the edge of the cloud, which, in combination with re-thermalization, reduces the temperature of the cloud and leads to condensate formation.

We investigate condensate formation further by plotting gas density profile snapshots in Fig.\ \ref{fig:densProf} at three different times during the simulation with $x_s=68.0\:\mu$m.
The lower panels show cross-sections of the thermal cloud density through the $y=0$ plane, whereas the upper panels show the full density, including any condensate, integrated along the $y$ direction. At the start of the simulation [Fig.\ \ref{fig:densProf}(a)], we see the thermal distribution expected of a gas above $T_c$. However, by $t=0.6\:$s [Fig.\ \ref{fig:densProf}(b)], a bimodal distribution has formed, suggesting the presence of a condensate, and a temperature below $T_c$. In the lower panels, a small dip in the central thermal density emerges, as the condensate mean-field potential forces thermal atoms from the center of the trap. By $t=2.25\:$s [Fig.\ \ref{fig:densProf}(c)], the condensate has grown and the thermal cloud has shrunk. The central dip in the thermal cloud has become more pronounced, giving rise to two density ``shoulders''. It should be noted that there is no rescaling of the vertical density scale in the upper panels, revealing the process does more than simply remove thermal atoms; $C_{12}$ collisions ensure that significant numbers of atoms are transferred to the condensate through re-thermalization. The ZNG method provides access to the condensate number through integration of the condensate wavefunction and returns $N_c \approx 11000$ for the $x_s=68.0\:\mu$m simulation and $N_c \approx 2540$ for the $x_s = 30.0\:\mu$m case \footnote{The condensate number was obtained by averaging $N_c$ between 1.5\:s and 3\:s}. Determination of experimental condensate fractions at such low atom numbers is extremely error prone, but may be estimated using absorption imaging and bimodal fitting. Obtained values of $N_c \approx 9000$ and $N_c \approx 3000$ for $x_s=68.0\:\mu$m and $x_s=30.0\:\mu$m respectively, at least reveal no serious inconsistencies.
\subsubsection{Distance Series}
\begin{figure}
\includegraphics{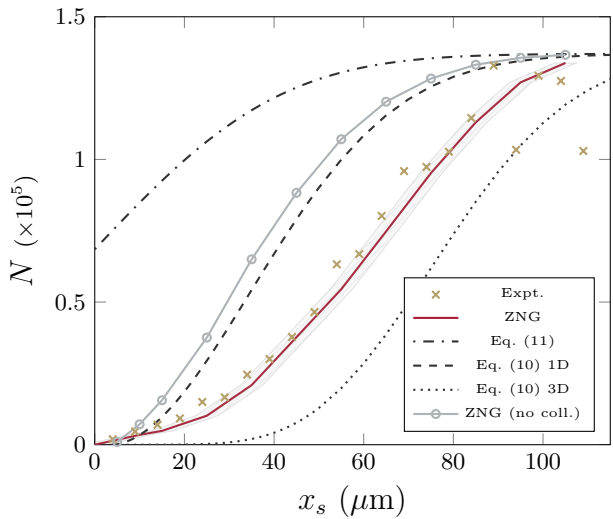}
\caption{(Color online) 
Total atom number, $N$, for a cloud held for $600\:$ms at a surface for varying trap-surface separations as measured experimentally (crosses) or simulated (solid curves). The main red solid curve corresponds to a ZNG simulation with collisions, the gray solid curve with open circles to a ZNG simulation without collisions ($C_{12} = C_{22}= 0$). The gray area around the red solid curve represents the error bounds assuming the surface position is shifted by $\pm 2.5\:\mu$m. The black dashed and dotted curves come from the classical model of Eq.~(\ref{eq:eta}) applied respectively in 1D and 3D (see text). The black dot-dash curve is an error function (Eq.\ (\ref{eq:errorFunc})) corresponding to the limit of very rapid transport.}
\label{fig:distance}
\end{figure}

In addition to the time-series results plotted in Fig.\ \ref{fig:lossRates}, plots of atom number as a function of distance from the surface, $x_s$, are of particular interest to experiments, revealing estimates of the cloud temperature and approximate surface position.  Figure \ref{fig:distance} shows the remaining atom number against $x_s$, each point corresponding to a single time-series simulation. These results were done at a slightly different temperature of 115\:nK with an initial atom number of $1.37 \times 10^5$ and a shorter hold-time of $0.6\:$s. Trap parameters are the same as for Fig.\ \ref{fig:lossRates}.

The red curve in Fig. \ref{fig:distance} is for a ZNG simulation with collisions and the gold cross points are from experiment. The gray curve with open circles is the same as for the ZNG simulation but without collisions ($C_{12} = C_{22} = 0$). We see again that collisions make an important contribution to the atom losses, causing greater losses further away from the surface and influencing the functional form of the curve. The collisionless simulations deviate strongly from the experimental data, as observed in our time-series simulations.

To further analyze these results, we present simplified atom loss calculations employing a classical model \cite{Lin2004}, which neglects Bose enhancement, re-thermalization and other dynamical effects within the cloud. It is based on the total energy distribution $n(E)\sim D(E) \exp(-E/k_B T)$ of thermal atoms, which is given by the density of energy states $D(E)$ and the corresponding Boltzmann factor with $k_B$ the Boltzmann constant and $T$ the temperature. For traps of limited depths, $\Delta V$, the fraction of atoms remaining in the trap can then be approximated by
\begin{equation}
\label{eq:partNumb}
\frac{N(\Delta V)}{N_0} = \frac{\int_0^{\Delta V}D(E)e^{-E/k_B T}dE}{\int_0^{\infty}D(E)e^{-E/k_B T}dE}.
\end{equation}
For a harmonic trap of dimensionality $d$, $D(E)\sim E^{d-1}$. Introducing the dimensionless parameter $\eta=\Delta V/k_B T$, Eq.\ (\ref{eq:partNumb}) becomes
\begin{align}
\label{eq:eta}
\frac{N(\eta)}{N_0} = \left\{
\begin{array}{ll}
1 - e^{-\eta} & \mbox{for}\;d=1 \vspace{0.15cm}\\
1 - (1+\eta)e^{-\eta} & \mbox{for}\;d=2 \vspace{0.15cm}\\
1 - (1+\eta + \frac{1}{2}\eta^2)e^{-\eta} & \mbox{for}\;d=3.
\end{array}
\right.
\end{align}
Starting from Eq.\ (\ref{eq:potential}) we can now calculate the trap depth $\eta(x_s)$ as a function of the trap-surface separation, which, together with Eq.\ (\ref{eq:eta}), allow us to model the number of remaining atoms $N(\eta(x_s))$ in a trap close to the surface.

The black dashed curve and black dotted curve in Fig.\ \ref{fig:distance} show the simple classical model results for $d=1$ and $d=3$, respectively. As this model neglects collisions, we compare it with the ZNG simulation without collisions [gray solid curve with open circles] and find approximate agreement with the $d=1$ curve. This is expected as the process of collisionless surface evaporation corresponds to a one dimensional loss channel; the small shift between the two curves is most likely due to the Casimir-Polder induced change in the density of states and the Bose enhancement in the initial state, which are both neglected in the classical model. Compared with the non-collisional curves, the full ZNG simulations show much larger loss-rates, which are consistent with the experiments. This is due to atomic collisions, causing an energy re-distribution between different directions. Atoms with large kinetic energy perpendicular to the surface normal can thus be scattered towards the surface and out of the trap. This re-distribution can be mimicked in the classical model by increasing the dimensionality of the loss channel. In the limit of $d=3$, the trap depth is effectively reduced in all directions, similar to radio-frequency (RF) evaporation. The corresponding curve [black dotted curve], however, has to be seen as an upper limit for the expected losses, as it lacks a proper description of re-thermalization effects; re-thermalization causes the cloud to shrink, thereby reducing the loss rates. The experimental results and the ZNG simulations lie between the 1D and 3D limits as expected. 

If the cloud is brought to and from the surface very rapidly, such that the period in which losses may be observed, $t_{\text{overlap}}$, is much shorter than the relevant trap period and thermalization time \footnote{Note that $t_{\text{overlap}}$ is always larger than the hold time since it includes losses during transport.}, it is possible to fit the loss curves with a complementary error function, where the remaining atom number, $N$, is given by an integral over a truncated Gaussian function
\begin{align}
\label{eq:errorFunc}
N&=N_0\cdot\int_{0}^\infty\sqrt{\frac{\alpha}{\pi}}e^{-\alpha(x-x_s)^2}dx.
\end{align} 
Here $N_0$ is the initial atom number and $\alpha=m \omega_x^2/2k_b T$.
With the help of this error function, the cloud temperature and surface position can be estimated, as has been done in previous studies \cite{gierling2011cold, Marchant2011}.
This limiting case, corresponding to $t_{\text{overlap}}=0$, is plotted for the experimental parameters ($T=115\:$nK) in Fig.\ \ref{fig:distance} [black dot-dash curve]. We note that this curve deviates significantly from all other models (which account for $t_{\text{overlap}}>0$) as the error function describes the case of swift transport that does not induce in-trap sloshing. The curve of Eq. (\ref{eq:errorFunc}) thus sets an upper limit to the number of atoms left at any distance.

Figure 4 indicates that in contrast to radio-frequency evaporative cooling, there are additional atom loss regimes when a surface is involved. At one end there is a ``fast limit'': if the cloud transport to and from the surface is very rapid- the surface acts as a pure spatial cutoff and the atom losses may be described with the error function [black dot-dash curve]. In this regime collisions and re-thermalization play no role. At the other end, there is a ``slow limit'' when $t_{\text{overlap}}$ is much greater than the thermalization time: in this case the surface acts much like a pure energy cutoff, as in the case of RF cooling. In between these limits, there is a further regime in which the re-thermalization is negligible (no collisions) but $t_{\text{overlap}}$ is greater than the trap period. In this case we have a 1D energy cutoff and the results may be described by collisionless models, such as Eq.\ (\ref{eq:eta}) [black dashed curve] (we note that the ZNG simulations with no collisions agree well with such results [gray curve with open circles]).  However, comparison of the collisionless ZNG results [gray curve with open circles] with both the experimental results [gold crosses] and the full ZNG simulation [red solid curve] shows the importance of including the full collisional dynamics of the system if thermalization becomes important, but the system is not yet in the slow limit.


\subsection{Optimizing condensate formation}

In this section we discuss how to improve condensate formation using surface evaporative cooling. Because the condensate atom number is influenced by many factors such as trap frequencies, transport velocity of the cloud, distance to the surface, initial atom number and initial temperature, a full exploration of the parameter space is not possible, given that the simulations can last on the order of tens of hours, even in parallel. We therefore focus our analysis here on a simplified system of an isotropic trap and a constant transport velocity, which should at least provide some basic guidance on how to obtain large condensate fractions and condensate numbers.

We consider a trap with $10^5$ $^{87}$Rb atoms and investigate the formation of a condensate for four different isotropic trap frequencies $\omega_1 =2\pi \times 40\:\text{rads}\:\text{s}^{-1}$, $\omega_2 =2\pi \times 80\:\text{rads}\:\text{s}^{-1}$, $\omega_3 =2\pi \times 120\:\text{rads}\:\text{s}^{-1}$ and $\omega_4 =2\pi \times 160\:\text{rads}\:\text{s}^{-1}$. The temperature of the initial equilibrium states is equal to the critical temperature $T=T_c$ and the cloud is prepared at $2.2\: W_l$ away from the surface, where $W_l=\sqrt{2k_bT/m\omega_n^2}$ is the initial thermal width of the cloud; this leads to different starting positions for each trap frequency.
Fixing the total evolution time at $0.75\:$s while keeping the same transport speed $0.1\:$mm$\text{s}^{-1}$, yields a variable hold time in each case. These hold times always exceed $0.4\:$s, which allows for sufficient equilibration. 

\begin{figure}
\includegraphics{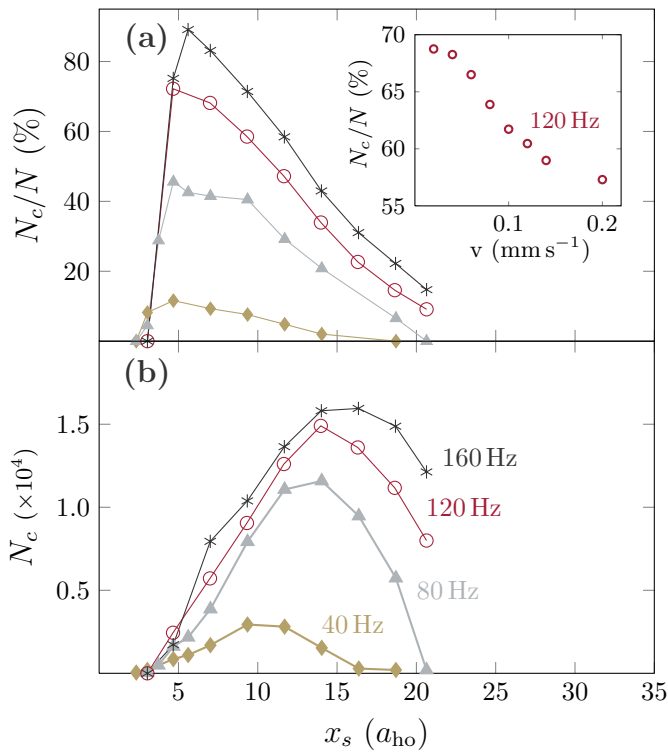}
\caption{(Color online) Condensate fraction $N/N_c$ (a) and condensate atom number (b) for different hold points, $x_s$, in the case of an isotropic trap, in units of the harmonic oscillator length $a_{\text{ho}}$. We show the results for four different trap frequencies $\omega_1 =2\pi \times 40\:\text{rads}\:\text{s}^{-1}$ (gold diamond points), $\omega_2 =2\pi \times 80\:\text{rads}\:\text{s}^{-1}$ (gray triangles), $\omega_3 =2\pi \times 120\:\text{rads}\:\text{s}^{-1}$ (open red circles) and $\omega_4 =2\pi \times 160\:\text{rads}\:\text{s}^{-1}$ (black stars). Condensate fractions are highest for these parameters between $4\:a_{\text{ho}}$ and $7.5\:a_{\text{ho}}$, whereas the condensate atom number has a maximum between 10.0\:$a_{\text{ho}}$ and 15.0\:$a_{\text{ho}}$. The inset in (a) shows condensate fractions plotted against transport velocity for a trap with frequency  $\omega_3 =2\pi \times 120\:\text{rads}\:\text{s}^{-1}$.}
\label{fig:condFracDist}
\end{figure}

Figure \ref{fig:condFracDist} shows condensate fraction (a) and condensate number (b) plotted against hold position, $x_s$, which is defined here in terms of the harmonic oscillator length $a_{\text{ho}}=\sqrt{\hbar/m\omega_n}$ to aid comparison.
Gold diamond-shaped points show the condensate fraction for $\omega_1$, gray triangles for $\omega_2$, open red circles for $\omega_3$ and black stars for $\omega_4$. 
In all four cases the condensate fraction is highest between $4\:a_{\text{ho}}$ and $7.5\:a_{\text{ho}}$. Although we observe condensate fractions of up to 90\% in that region, the corresponding condensate atom numbers are relatively small, as shown in Fig.\ \ref{fig:condFracDist}(b). This is due to the fact that the condensate is already in contact with the surface at these distances, leading to the loss of ground state atoms. Higher condensate atom numbers are achieved further away from the surface, with the optimal distance between around 10$\:a_{\text{ho}}$ and 15$\:a_{\text{ho}}$ for the parameters considered here. In order to maximize the condensate fraction and minimize the remaining thermal atoms at the same time, a holding distance $x_s \sim$ 10$\:a_{\text{ho}}$ appears to be a good compromise. As expected, higher trap frequencies increase the atom density and hence improve re-thermalization properties leading to faster formation of larger condensates.  

We now briefly consider the influence of the transport velocity $v$, at which the cloud is brought to the surface, on condensate formation [Fig.\ \ref{fig:condFracDist}(a), inset]. We see that if the transport velocity is reduced below 0.1\:mm\:s$^{-1}$ for the case with $\omega_3=120$\:Hz and hold position $1\:W_l$ ($9.3\:a_{\text{ho}}$), the condensate fraction increases from $\sim 60\%$ to $\sim 70\%$. However, it saturates when going to velocities $\lesssim 20\:\mu$m\:s$^{-1}$. In terms of condensate fraction there is, therefore, little to be gained through lower approach speeds. We have checked that for faster transport speeds of 1\:mm\:s$^{-1}$, the best hold position is roughly in the same place. 

The results in Fig.\ \ref{fig:condFracDist} are for a very cold cloud at $T_c$. In order to mimic a more realistic starting state when atoms are loaded into a chip trap, we have performed further simulations for $\omega_4 =2\pi \times 160\:\text{rads}\:\text{s}^{-1}$, but with a starting temperature of 2\:$\mu$K and $2\:\times10^{6}$ atoms. We use the hold position of 10\:$a_{\text{ho}}$, as suggested by the results in Fig.\ \ref{fig:condFracDist}. Due to the higher atom number, condensates of  $\sim 90$\% purity and $\sim 100,000$ atoms were observed.
It is interesting to note that our ZNG method remains at least qualitatively correct even for this unusually warm starting state. It suggests that the primary limitations of the method are the simulation run time and the $s$-wave scattering approximation, which is valid up to approximately 100\:$\mu$K \cite{Wade2011}.
%

\section{CONCLUSIONS}
In conclusion, we have studied the evaporative cooling of cold atom clouds
at surfaces, finding that the ZNG method provides a satisfactory
description of the physics provided a full numerical implementation of the
collisions is included. We have seen that there are multiple atom
loss regimes; only very fast transport of the atom cloud permits an
analytic description of atom losses under normal circumstances, a fact
which has important implications for surface calibration in experiments.
Finally, we suggest that at least for cold starting temperatures, the
purest condensates are achieved by bringing the cloud to a separation from
the surface of around 5 harmonic oscillator units. The biggest condensates,
however, were achieved for separations of 10 – 15 oscillator units.

For a complete cooling scheme of $^{87}\text{Rb}$ atoms in a chip trap, we suggest starting by bringing the cloud rapidly 
to a hold position $x_s\approx3.0\:{W_l}$ while avoiding in-trap sloshing. At this point there is negligible overlap between the cloud and the surface.
The trap should then be moved with a velocity $v\lesssim 0.1\:$mm$\text{s}^{-1}$
to a final hold position at about $10\:a_{\text{ho}}$. 
In this way a large and relatively pure condensate can be achieved even for
low trap frequencies in a time which is comparable with that for conventional
RF cooling.

We end with a few general remarks and a discussion of how surface
cooling compares with traditional radio frequency cooling. There were
initial concerns that surface cooling might be less efficient than RF
cooling, being a one-dimensional cooling. However, Fig.\ \ref{fig:distance} suggests that cooling for this system lies between 1D and 3D cooling, allowing efficient formation of BECs. Condensate formation times do not appear to compare poorly with those reported for RF cooling \cite{Harber2003}.
In addition, surface cooling may convey certain advantages: the Casimir-Polder
potential leads to a sharper energy ``knife'' and the positioning
accuracy limits of atom chips ($<2\:$nm \cite{gierling2011cold}) might allow competitive
control of the energy barrier height, without the need for a signal
generator. Furthermore, the ability to use a long cloud axis for cooling
might offer greater control with respect to in-trap oscillations.

\vspace{2cm}

\begin{acknowledgments}
 We gratefully acknowledge financial support from the Carl-Zeiss foundation, the BW-Stiftung ``Kompetenz\-netz funktionelle Nanostrukturen'', the Deutsche Forschungsgemeinschaft through SFB TRR21 and the UK EPSRC (grant no.\ EP/I019413/1). We also thank Eugene Zaremba for helpful discussions and BW-Grid computing resources. 
\end{acknowledgments}

\bibliography{bibFile}
\end{document}